\documentclass[final,5p,times, twocolumn]{elsarticle}
\usepackage{booktabs}
\usepackage{lineno}
\usepackage{amsmath}
\usepackage{amssymb}
\usepackage{subfig}
\usepackage{ifpdf}
\usepackage{graphicx}
\journal{Nuclear Instrumentation and Methods A}
\begin{document}
\newcommand{\bbdk}  {0 \nu \beta \beta}
\newcommand{\iso}[2]{$^{#1}$#2}
\newcommand{\mc}[2]{\multicolumn{#1}{c}{#2}}


\title{Alpha Backgrounds for HPGe Detectors in \\ Neutrinoless Double-Beta
Decay Experiments}

\author[UW]{R.A.~Johnson\corref{corUW}\fnref{fCU}}
\ead{robj@npl.washington.edu}
\author[UW]{T.H.~Burritt}
\author[LANL]{S.R. Elliott}
\author[LANL]{V.M. Gehman\fnref{fLBNL}}
\author[SDU]{V.E. Guiseppe}
\author[UNC,TUNL,ORNL]{J.F. Wilkerson}
\cortext[corUW]{Corresponding author}
\fntext[fCU]{Currently at University of Colorado, Boulder, CO 80309}
\fntext[fLBNL]{Currently at Physics Division, Lawrence Berkeley National Laboratory, Berkeley, CA 94720}
\address[UW]{Center for Experimental Nuclear Physics and Astrophysics,
University of Washington, Seattle, WA 98195, USA}
\address[LANL]{P-23, Los Alamos National Laboratory, Los Alamos, NM, 87545, USA}
\address[SDU]{University of South Dakota, Vermillion, SD 57069, USA}
\address[UNC]{University of North Carolina, Chapel Hill, NC, 27599, USA}
\address[TUNL]{Triangle Universities Nuclear Laboratory, Durham, NC, 27708, USA}
\address[ORNL]{Oak Ridge National Laboratory, Oak Ridge, TN 3783, USA}

\date{\today}
\begin{abstract}

The {\sc Majorana} Experiment will use arrays of enriched HPGe detectors to
search for the neutrinoless double-beta decay of $^{76}$Ge. Such a decay, if
found, would show lepton-number violation and confirm the Majorana nature of the
neutrino.  Searches for such rare events are hindered by obscuring backgrounds
which must be understood and mitigated as much as possible.  A potentially
important background contribution to this and other double-beta decay
experiments could come from decays of alpha-emitting isotopes in the $^{232}$Th
and $^{238}$U decay chains on or near the surfaces of the detectors. An alpha
particle emitted external to an HPGe crystal can lose energy before entering the
active region of the detector, either in some external-bulk material or within
the dead region of the crystal. The measured energy of the event will only correspond to a
partial amount of the total kinetic energy of the alpha and might obscure the
signal from neutrinoless double-beta decay.  A test stand was built and
measurements were performed to quantitatively assess this background.  We
present results from these measurements and compare them to simulations using
Geant4.  These results are then used to measure the alpha backgrounds in an
underground detector {\em in situ}.  We also make estimates of surface
contamination tolerances for double-beta decay experiments using solid-state
detectors.

\end{abstract}


\maketitle

\section{Introduction}
\label{sec:Intro}

Observation of neutrinoless double-beta decay $(\bbdk)$ represents the best
chance for discovering the nature of the neutrino (Majorana or Dirac)
\cite{Ell02, Ell04,Eji05,Avi07}.  Experimental searches using high-purity
germanium (HPGe) detectors---enriched to 86\% \iso{76}{Ge}---have demonstrated
the most stringent half-life limits for this decay \mbox{($T_{1/2}^{0\nu} > 1.9
\times 10^{25}$ y \cite{Kla01})} to date.  Future experiments with tonne-years
of exposure should be able to reach half-life limits greater than $10^{27}$ y,
but only through concerted efforts in understanding and reducing backgrounds.
Current experiments using \iso{76}{Ge}, such as {\sc Majorana}\cite{Sch11,Phi11,Agu11}
and GERDA Phase II\cite{Sch05, Jan09}, will attempt to reach background levels of
$\sim$ 4 background events/tonne-year in the $0\nu\beta\beta$ region-of-interest
around 2039 keV to demonstrate the viability of future tonne-scale
experiments\footnote{A background of 4 events/tonne-year in the {\sc Majorana}
and GERDA experiments will scale to $\sim$ 1 event/tonne-year in a tonne-sized
experiment.}.  Achieving these background goals will require ultra-clean
materials, sufficient shielding, sophisticated background-rejection techniques,
and a deep-underground setting.  

The signature of $\bbdk$ is a peak at the Q-value of the decay, corresponding to
the sum of the kinetic energies of the two emitted electrons.  Alpha particles,
emitted from decays in the \iso{232}{Th} and \iso{238}{U} natural decay chains
with energies of 3.9--8.8 MeV, can lose kinetic energy before entering the
active region of an HPGe crystal.  These degraded alphas can result in a
continuum of events, obscuring a possible signal from $0\nu\beta\beta$.

\begin{figure}[h]
 \includegraphics[width=\linewidth]{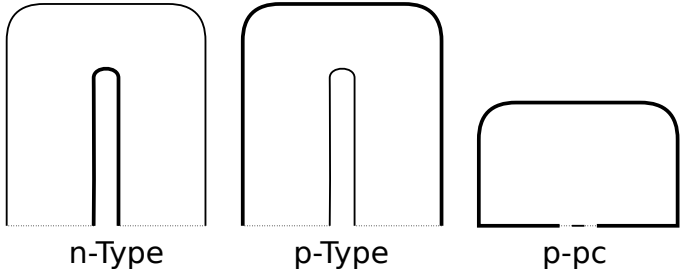}
 \caption{\label{fig:DetectorTypes} Cross-section diagrams of different HPGe
 detectors.  The thin outline represents the $p^{+}$ (thin) dead layer and the
 thick outline represents the $n^{+}$ dead layer.  The n- and p-type detectors
 are of the common semi-coax design.  The p-type point contact detector has far
 less thin surface than either the p- or n-types.  The area on the bottom of the
 crystals (dashed line) represents the passivated surface that insulates the
 $n^+$ and $p^+$ contacts.  
 }
\end{figure}

HPGe diode detectors have $n^+$ and $p^+$ ohmic contacts for bias and signal
connection.  The $n^+$ layer, created with diffused lithium ions, is typically
$0.5-1$ mm in depth.  The $p^+$ layer is created by implanting boron ions,
resulting in a dead layer on the order of $\sim 0.5 \mu$m.  These form regions
within the detector that are insensitive to ionizing radiation, where energy
lost by a particle will not be registered.  Because alphas cannot traverse the
thick $n^+$ layer, a detector's susceptibility to alphas is dependent upon the
amount of surface with a $p^+$ contact.  HPGe detectors are characterized by
their dopants, being either n-type or p-type. The dopants then determine the
type of ohmic contacts for the crystal; n-type detectors have $n^+$ central
contacts while p-types have $p^+$.  The outer surface contacts are reversed for
each detector type.  The diagram in Fig.  \ref{fig:DetectorTypes} shows a
depiction of these detector types with exaggerated thicknesses.  The thin outer
contact means that n-type detectors have more alpha-susceptible surface area
than p-types.  A p-type with a point
contact (p-pc, \cite{Bar07}) will be used in the {\sc Majorana
Demonstrator}\cite{Phi11} and Phase II of GERDA \cite{Ago11}.
  The region between the $n^+$ and $p^+$ layers is
typically passivated and is characterized by incomplete and hard to predict charge
collection (shown as dashed in the figure).  The response of a detector to
alphas incident on this surface is still an unknown. The p-pc detectors
used in the {\sc Majorana Demonstrator } will have a minimal amount of this type
of surface.

Broadly speaking, alpha backgrounds for $\bbdk$ experiments with HPGe detectors
fall into two categories, depending on where the degradation in kinetic energy
occurs: surface alphas and external-bulk alphas.  

Surface-type alpha events are characterized by decays at the surface of the
crystal, with the alpha losing kinetic energy within the dead region of the
crystal.  Only alphas that enter the crystal at extremely shallow incidence
angles (with respect to the normal of the crystal surface) will lose an
appreciable amount of energy within this thin dead region; alphas entering the
crystal at or near normal to the surface will lose a negligible amount of energy before entering
the active region of the crystal, resulting in minimal energy loss and a peak
structure when measured by the detector.  The classic example of this background
arises from exposure to \iso{222}{Rn}.  The decay of \iso{222}{Rn} (3.8 days)
and its subsequent daughters \iso{218}{Po}, \iso{214}{Pb}, \iso{214}{Bi},
\iso{214}{Po} eventually lead to \iso{210}{Pb}.  Along the way, these daughters
of \iso{222}{Rn} can implant onto surfaces\cite{Leu06}.  \iso{210}{Pb} decays to
\iso{210}{Bi}, and then \iso{210}{Po}, which emits a 5.3 MeV alpha upon its
decay.  The relatively-long half-life of \iso{210}{Pb} (22 years) means that
even a brief exposure of a detector or its surroundings can lead to a steady
supply of 5.3 MeV alpha backgrounds\cite{Sto05, Leu06}.

The other category of alpha backgrounds, external-bulk alphas, originate in bulk
materials external to the crystal, { \em e.g.}\ a contact pin or detector mount.
Energy loss of the alpha occurs in this bulk material before it hits the
detector surface and depends on the amount of external material the alpha
travels through.  The result is a broad continuum of events with no peak
structure.

This paper presents work to quantify the role of alpha backgrounds on $\bbdk$
experiments using HPGe detectors.  in Sec. \ref{sec:TestStand}
a test stand that was designed and built to
study alpha backgrounds is described.  Also in this section a model of the
response of an HPGe detector to alpha decays is discussed and applied to
the test stand data.  In Sec. \ref{sec:InSitu} this model is applied
to data taken with an {\em in situ} detector located underground.
These results are discussed in the context of backgrounds to
$\bbdk$ experiments using HPGe detectors in Sec. \ref{sec:Impact}.

\section{Background Test Stand}
\label{sec:TestStand}


An n-type, semi-coax HPGe detector manufactured by ORTEC\cite{ORTEC} was
modified to facilitate measurement of an alpha source on the surface of an HPGe
crystal.  The original detector was enclosed within an aluminum outer vacuum
cryostat that sealed against the base of the detector canister.  The base is
mounted securely to the cold finger and contains all electrical feedthroughs.
The inside of the canister contains the upper cold finger and cold plate,
front-end electronics for the preamplifier, and the crystal.  The crystal mount
consists of a smaller aluminum cup attached to the cold finger with the HPGe
crystal  affixed inside this mounting cup.  A thin sheet of aluminized mylar
that originally covered the crystal's front face was replaced with a crystal cap
fabricated from aluminum.  The external vacuum enclosure---sealed against the
detector base with an O-ring---was replaced with a larger, custom cryostat
enclosure for ease of source transfer and for external-source manipulation.  The
test stand was operated in two separate modes to study both surface-type and
bulk-type alpha backgrounds.

\begin{figure}[h]
\includegraphics[width=\linewidth]{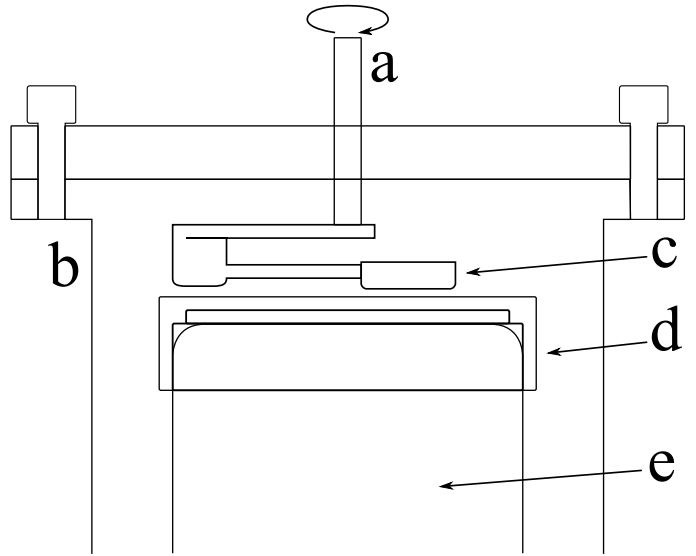}
\caption{
Side-view of top portion of modified detector cryostat enclosure.  A rotational
feedthrough (a) is built into the the modified cryostat enclosure (b).  The
feedthrough rotates a source (c) in a circle above the collimation plate (d).
The collimation plate (Fig. \ref{fig:Collimation}) rests on the crystal mounting
cup above the HPGe crystal (e) and provides a collimated source of alphas from
the source to illuminate the crystal face.  } 
\label{fig:SantaSchematicI} 
\end{figure}

\begin{figure}[h]
\centering
\subfloat[Top-down view]{
  \includegraphics[width=0.9\linewidth]{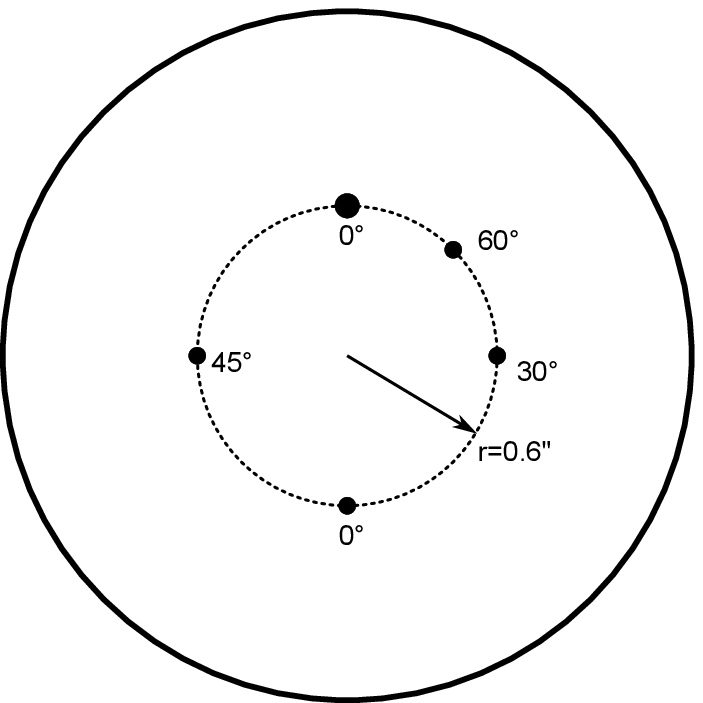}
}

\subfloat[Side view]{
  \includegraphics[width=0.9\linewidth]{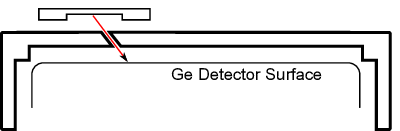}
}
\caption{\label{fig:Collimation} 
The plate that sits just atop the HPGe crystal has collimation holes, drilled at
$0^{\circ}$, $30^{\circ}$, $45^{\circ}$, and $60^{\circ}$ with respect to the
normal of the plate, and all situated at a radius of 0.6'' from the center of
the plate (a). The alpha source is constrained to rotate above this circle, allowing
it to shine through any one of these holes via an external 
rotational feedthrough built into the detector
cryostat. The side view (b) shows the source and alpha shine through a
representative hole onto the surface of the HPGe detector. }
\end{figure}

\subsection{Surface Background Model}
\label{sec:SurfaceBackgrounds}

\begin{table}[h]
\centering
\caption{\label{tab:241AmSource}
The alpha decay of \iso{241}{Am} has five prominent alphas and two
prominent gamma lines from the decay to \iso{237}{Np} (22.35 keV at 2.27\% and 59.54 keV at 35.9\% branching ratio). The two
gammas only accompany the main alpha peak (c), leading to alpha-gamma pileup
that contributes more counts to peaks d and e. Values of energy and branching ratio taken from
\cite{Bas06}.}

\begin{tabular}{lr@{.}lr@{.}l}
\toprule
Peak  & \multicolumn{2}{c}{Centroid (keV)} & \multicolumn{2}{c}{B.R.(\%)}
\\ \midrule
a   & 5388&0  &       1&66   \\
b   & 5442&8  &      13&1    \\
c   & 5485&56 &      84&8   \\
d   & 5511&5  &       0&225 \\
e   & 5544&5  &       0&37  \\
\bottomrule
\end{tabular}
\fntext[pileup]{Also includes counts from alpha-gamma pileup.}
\end{table}

\begin{figure}[h]
\includegraphics[width=\linewidth]{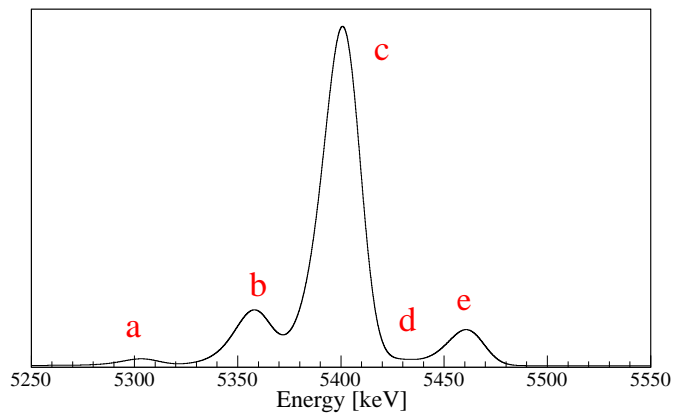}
\caption{\label{fig:Hole0} Representative energy spectrum (from simulation)  of a collimated
$^{241}$Am source.  The 5 peaks, labeled a-e, correspond to the peaks in Table
\ref{tab:241AmSource}.  The centroids of the peaks are lower in energy than the
centroids from Table \ref{tab:241AmSource} due to energy loss in the dead region
of the detector, as described in the text.}
\end{figure}

For studying surface-type backgrounds, the test stand (see schematic, Fig.
\ref{fig:SantaSchematicI}) employed a rotational feedthrough (a) at the top of
the modified cryostat enclosure (b).  The feedthrough connects to an arm and
source holder (c) in which a windowless-alpha source is placed.  The source can
be rotated in a circle above the aluminum cap (d), through which holes were
drilled at different angles (Fig.  \ref{fig:Collimation}).  These angled holes
allowed alphas of different incidence angles (0$^{\circ}$, 30$^{\circ}$,
45$^{\circ}$, and 60$^{\circ}$ with respect to the normal of the surface) to
reach the crystal's (e) face.  The source used was a windowless $^{241}$Am alpha
source from Isotope Products with an activity of 161.9 Bq
\cite{IsotopeProducts}.  The decay of \iso{241}{Am}$\rightarrow$\iso{237}{Np}
(Q-value: 5637.82 keV) results in the emission of an alpha particle 100\% of the
time.  The five alpha branches with probabilities greater than 0.1\% are noted
in Table \ref{tab:241AmSource}.  There are two gamma emissions with
non-negligible branching ratios that occur with the 5485.56 keV alpha.  Given
the proximity of the source to the detector, a large fraction of these decays
will deposit both an alpha and a gamma in coincidence, and the energy spectrum
as measured by the HPGe detector will reflect this with some of the events from
the 5485 keV alpha going into a higher energy peak.   This structure can be seen
in Fig.  \ref{fig:Hole0}.

Data were collected using four different collimation holes, corresponding to
alphas striking the surface of the detector at incidence angles as described
above.  Figure \ref{fig:HoleData} shows calibrated energy spectra for all four
data sets.  

A charged-ion interaction model was constructed to characterize the detector's
response to surface alphas.  An alpha traversing a non-active region of length
$\Delta x$ will undergo a mean energy loss of

\begin{equation} 
\label{eq:DeltaE} 
\Delta E = \int_0^{\Delta x}
\frac{dE}{dx}\left(E \right) dx, 
\end{equation}

\noindent where the stopping power (itself a function of the alpha's energy) is
integrated over $\Delta x$.  A Bohr model of non-relativistic heavy particles on
thick absorbers \cite{PDG10,Sig06} was used to relate the energy straggling to
the expected width of a mono-energetic alpha beam incident on an absorber ({\em
i.e.} a non-active region).  An idealized energy spectrum (as measured by an
HPGe detector) of alphas of initial kinetic energy $E_0$ traveling through an
absorber of thickness $\Delta x$ would then be described by a gaussian with mean
$E_0 - \Delta E$ and width $\sigma$:

\begin{equation}
\label{eq:Width}
\sigma^2 =  4\pi N_{a}r^2_{e}(m_{e}c^2)^2\rho\frac{Z}{A}\Delta x.
\end{equation}

Here, $N_a$ is Avogadro's number, $r_e$ is the classical radius of the electron,
$m_e$ the mass of the electron, and $\rho$, $Z$, and $A$ the density, atomic
number, and atomic mass of the absorber material.  In practice there are two
further required modifications.  Nuclear quenching within the HPGe detector
results in energy losses that do not register with the ionization detector.
This results in a further energy offset, $\Delta E_{NQ}$, that must be
included.  Furthermore, HPGe detectors do not have 100\% efficiency for charge
collection, resulting in a low-energy tail.  For this reason, an
exponentialy-modified Gaussian (a Gaussian convolved with an exponential) is
used in the model to account for this asymmetry in the peak signal.
Incorporating these two additions, the model-predicted energy spectrum from a
mono-energetic alpha source traversing an absorber becomes

\begin{align}
\label{eq:ModelSignal}
&\mathcal{G}(E, \mu_0, \sigma) = \\ \notag
&\frac{1}{2\tau} 
\exp\left( \frac{\sigma^2}{2\tau^2} + \frac{E-\mu_{0}}{\tau}\right) 
\mathrm{Erfc}\left(\frac{E-\mu_0}{\sqrt{2}\sigma} + \frac{\sigma}{\sqrt{2}\tau} \right)
\end{align}

\noindent where $\mu_0 = \Delta E + \Delta E_{\text{NQ}}$, $\tau$ is the
exponential parameter, and $\sigma$ is defined as in Eq. \ref{eq:Width}.  

Tailoring Eq. \ref{eq:ModelSignal} to the \iso{241}{Am} source in
the test stand requires summing 5 such peaks and also taking into account the
background from cosmic ray-induced events.  The cosmic background is
well-described by a linear polynomial ($\mathcal{P}_1$) in the region around the
peak structure (5100-5600 keV).  The energy loss and straggling parameters,
$\Delta E$ and $\sigma$, depend on the path length through the dead layer
(depth $\mathcal{D}$) via $\Delta x = \mathcal{D} / \cos\theta$, where $\theta$
is the incidence angle of the alpha.  
The probability density function describing surface
alphas from \iso{241}{Am} impinging on an HPGe crystal surface at incident angle
$\theta$ in the test stand is then

\begin{equation}
R(E, \theta, \mathcal{D}) = \frac{N_{\textrm{Am}}}{N_{\textrm{total}}}\sum^5_{k=1} c_k \mathcal{G}(E, \mu_k,
\sigma) + \frac{N_{\textrm{Bg}}}{N_\textrm{total}}\mathcal{P}_1,
\end{equation}

\noindent where the coefficients $c_k$ represent the weights of the alpha peaks
in Table \ref{tab:241AmSource} and sum to 1.  

\subsection{Test Stand Operation}

The test stand was operated at Los
Alamos National Laboratory at 2200 meters above sea level.  The high
cosmic-event rate, coupled with the relatively weak source (161.9 Bq), forced
the size of the collimation holes to be larger than ideal.  The poorer
collimation power resulted in a larger range of incidence angles for a
particular hole than desired, and required a modified probability density
function composed of a weighted sum of single-angle p.d.f.s.  The final p.d.f.,
taking into account the spread of angles around the nominal collimation angle
$\hat{\theta}$, then becomes 

\begin{equation} 
R(E,\hat{\theta}, \mathcal{D}) = \sum_i w_i
R(E,\theta_i, \mathcal{D}).  
\end{equation} 

\begin{figure*}[t]
\centering
\subfloat[$0^{\circ}$ incidence.]{\label{fig:FitHole0}
\includegraphics[width=0.48\textwidth]{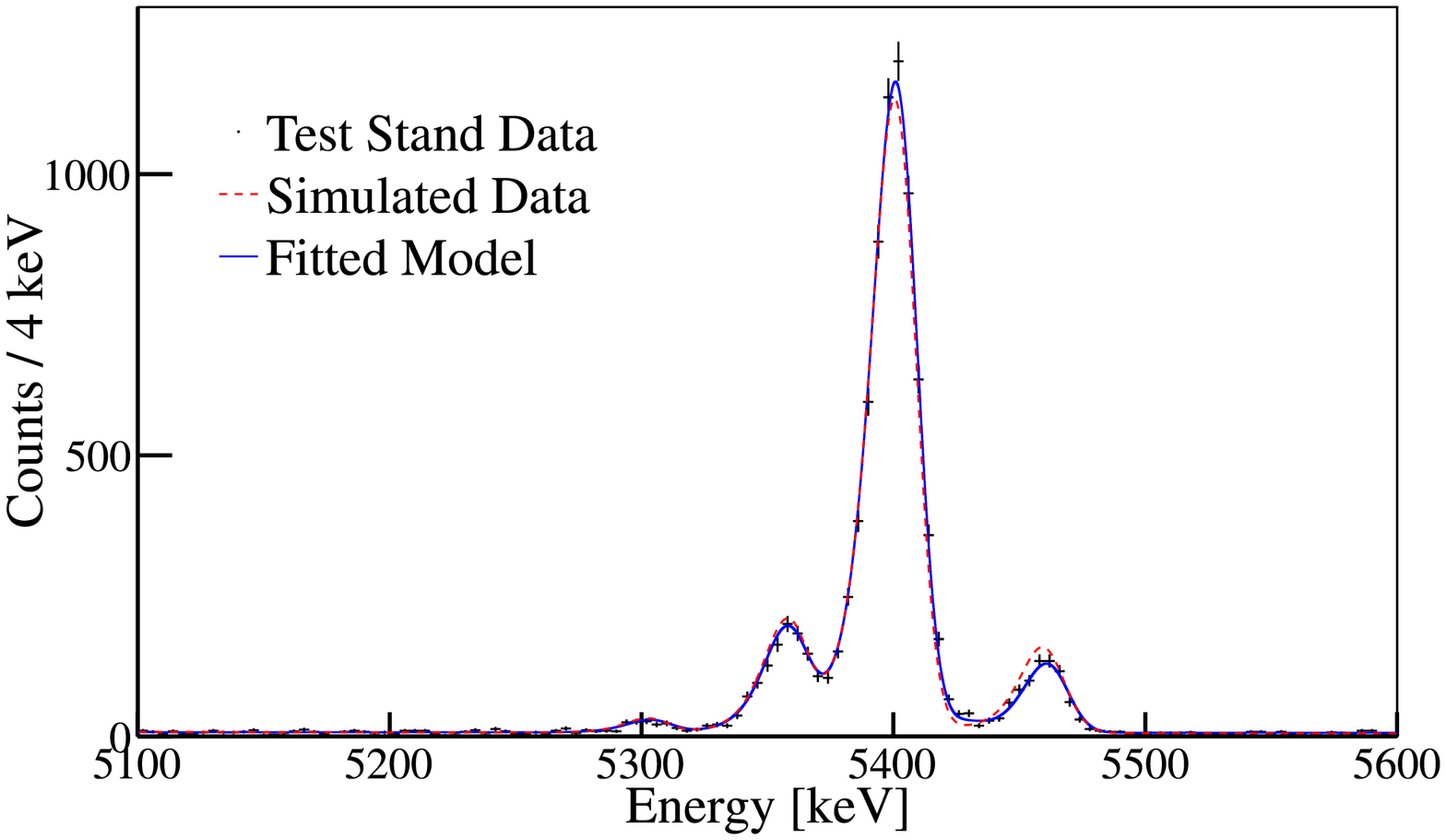}}
\subfloat[$30^{\circ}$ incidence.]{\label{fig:FitHole30}
\includegraphics[width=0.48\textwidth]{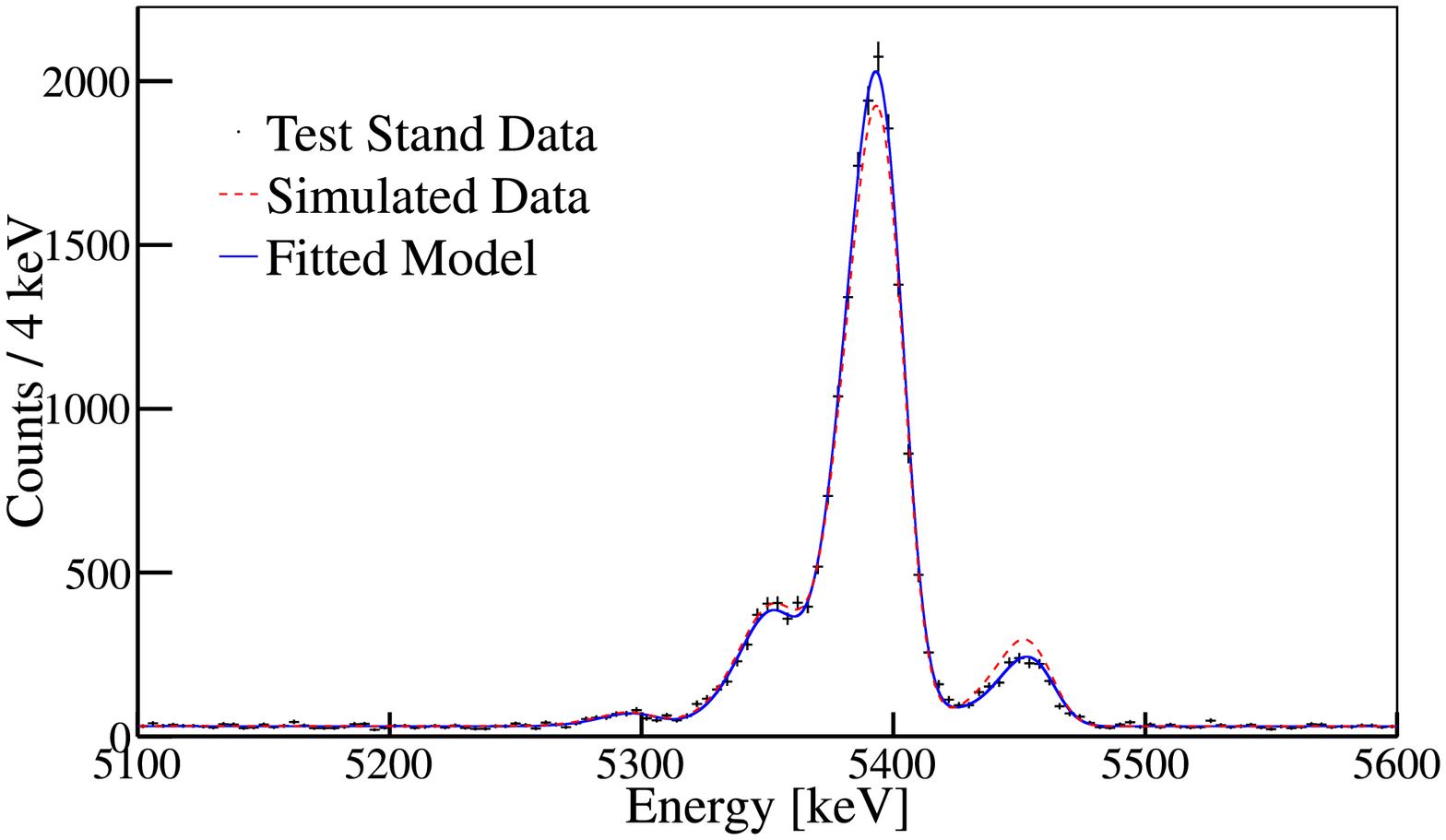}}

\subfloat[$45^{\circ}$ incidence.]{\label{fig:FitHole45}
\includegraphics[width=0.48\textwidth]{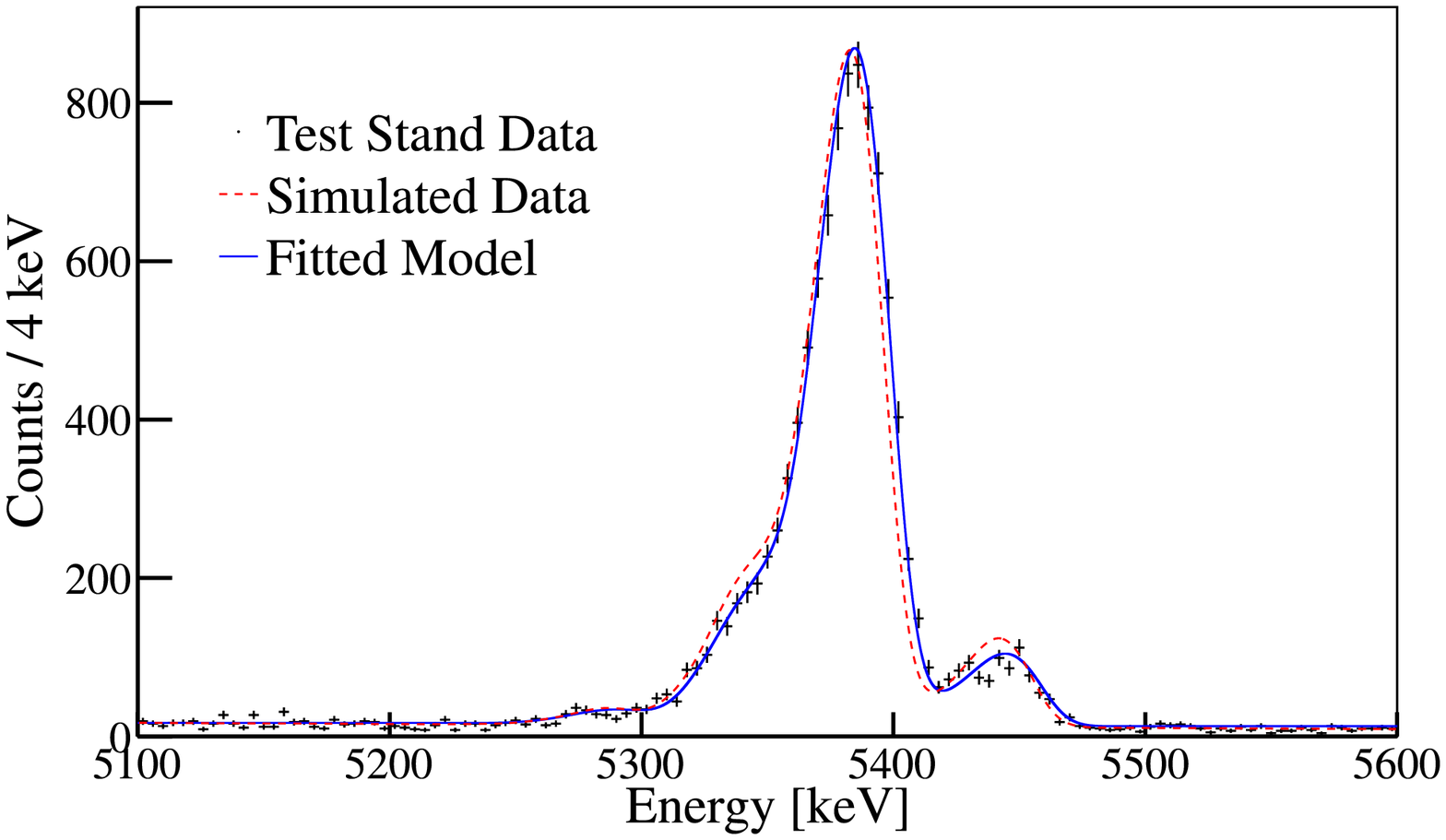}}
\subfloat[$60^{\circ}$ incidence.]{\label{fig:FitHole60}
\includegraphics[width=0.48\textwidth]{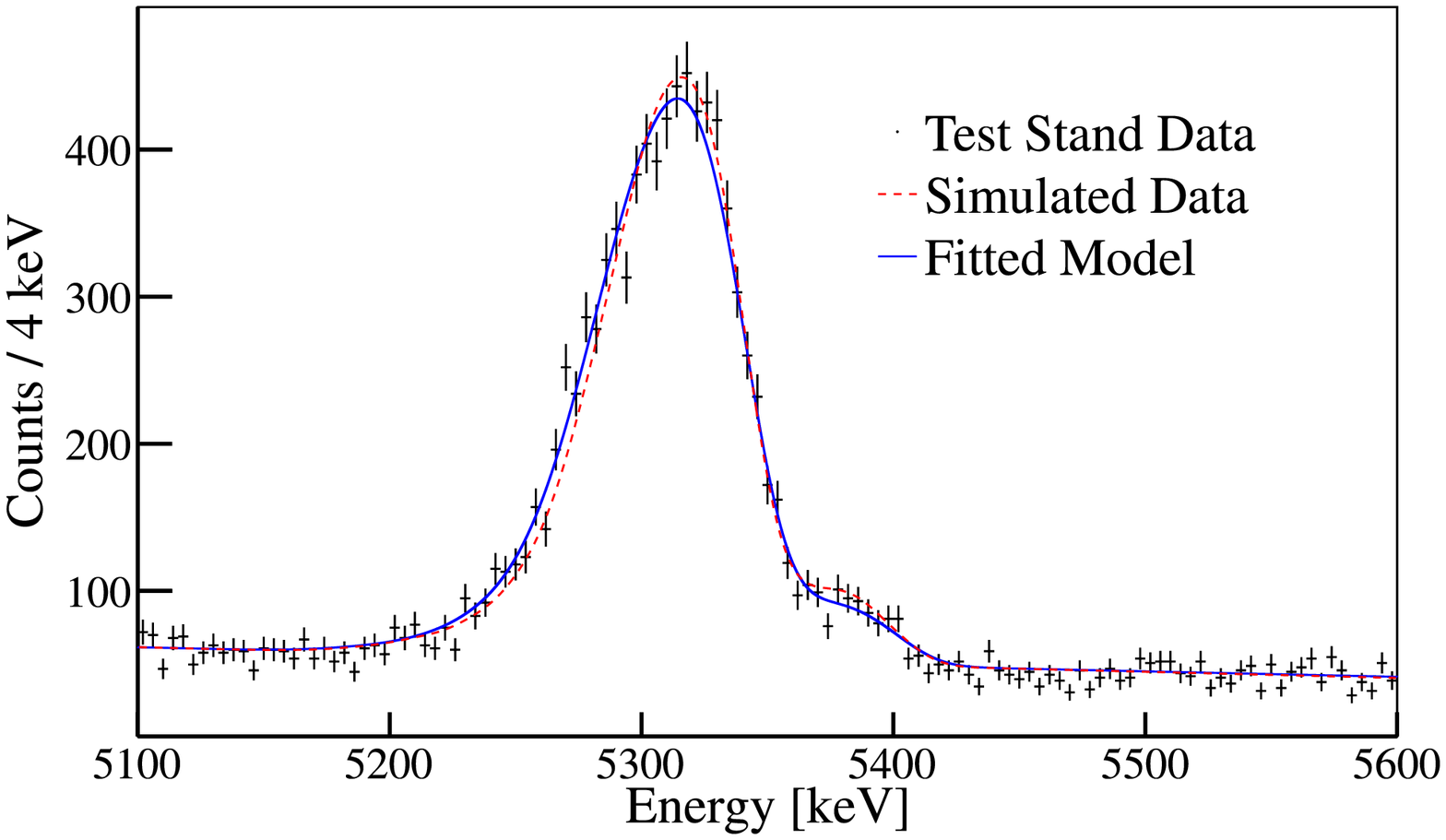}}
\caption[]{Surface-type alphas from \iso{241}{Am} at incidence angles  of $0^{\circ}$,
$30^{\circ}$, $45^{\circ}$, and $60^{\circ}$ with respect to the surface normal.
Also shown are the analytic-model fits (solid line) and simulated spectra from
Geant4 (dashed line, corrected via the convolution from Fig. \ref{fig:SimCorrection}).}
\label{fig:HoleData}
\end{figure*}

\noindent The weights $w_i$ for a given single incidence angle $\theta_i$ are
determined via a line-of-sight simulation that takes into account the
collimation angle, size, and the position of the source. It is important to note
that the peaks in Fig. \ref{fig:HoleData} are wider than they would be with a
perfect collimator.

The analytical model was fit to the test stand data using a maximum-likelihood
fit.  All four data sets (with nominal incidence angles of 0$^{\circ}$,
30$^{\circ}$, 45$^{\circ}$, 60$^{\circ}$) were fit simultaneously with the goal
of extracting the dead layer parameter $\mathcal{D}$.  The combined model for
the data sets with the data are shown in Fig. \ref{fig:HoleData}.  The value of
the dead layer, as extracted from the fits, is determined to be $0.307 \pm 0.005
\mu m$.  This is in agreement with the dead layer thickness as given by the
data sheet from ORTEC (0.3 $\mu$m, no stated uncertainty). 

\begin{figure}
\centering
\subfloat[Deficit in peak offset position ($\Delta E = \mu_{\text{anal}} -
\mu_{\text{sim}}$).]{\label{fig:SimulatedOffset}
\includegraphics[width=0.95\linewidth]{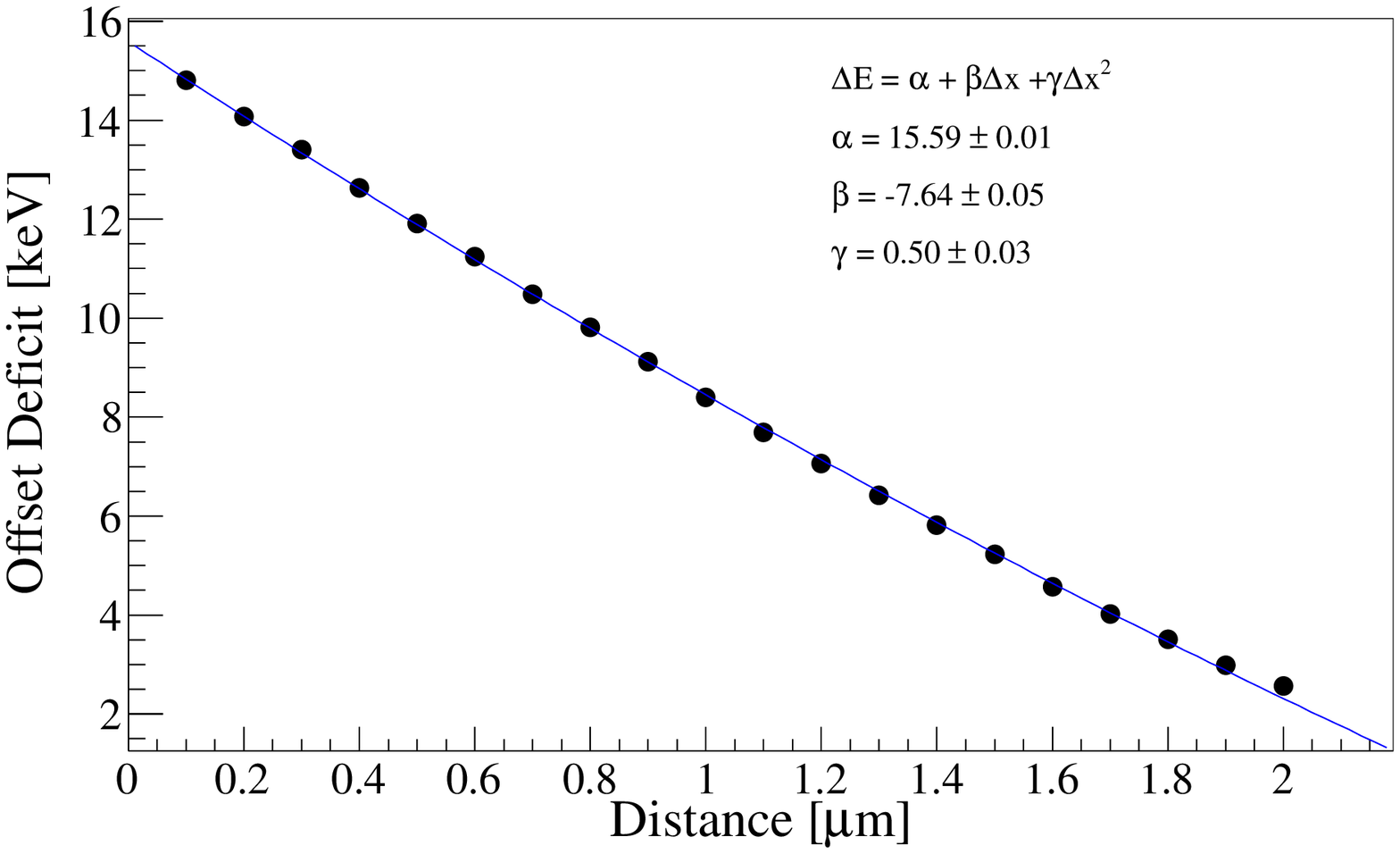}}

\subfloat[Differences in peak width between simulation and the model. A
parametric function of dead layer distance ($\sigma = \sqrt{\alpha + \beta
\delta x}$ is shown for the analytic model (red, dashed line) and is also fitted
to the simulated points (blue, solid line).]
{\label{fig:SimulatedSigma}
\includegraphics[width=0.95\linewidth]{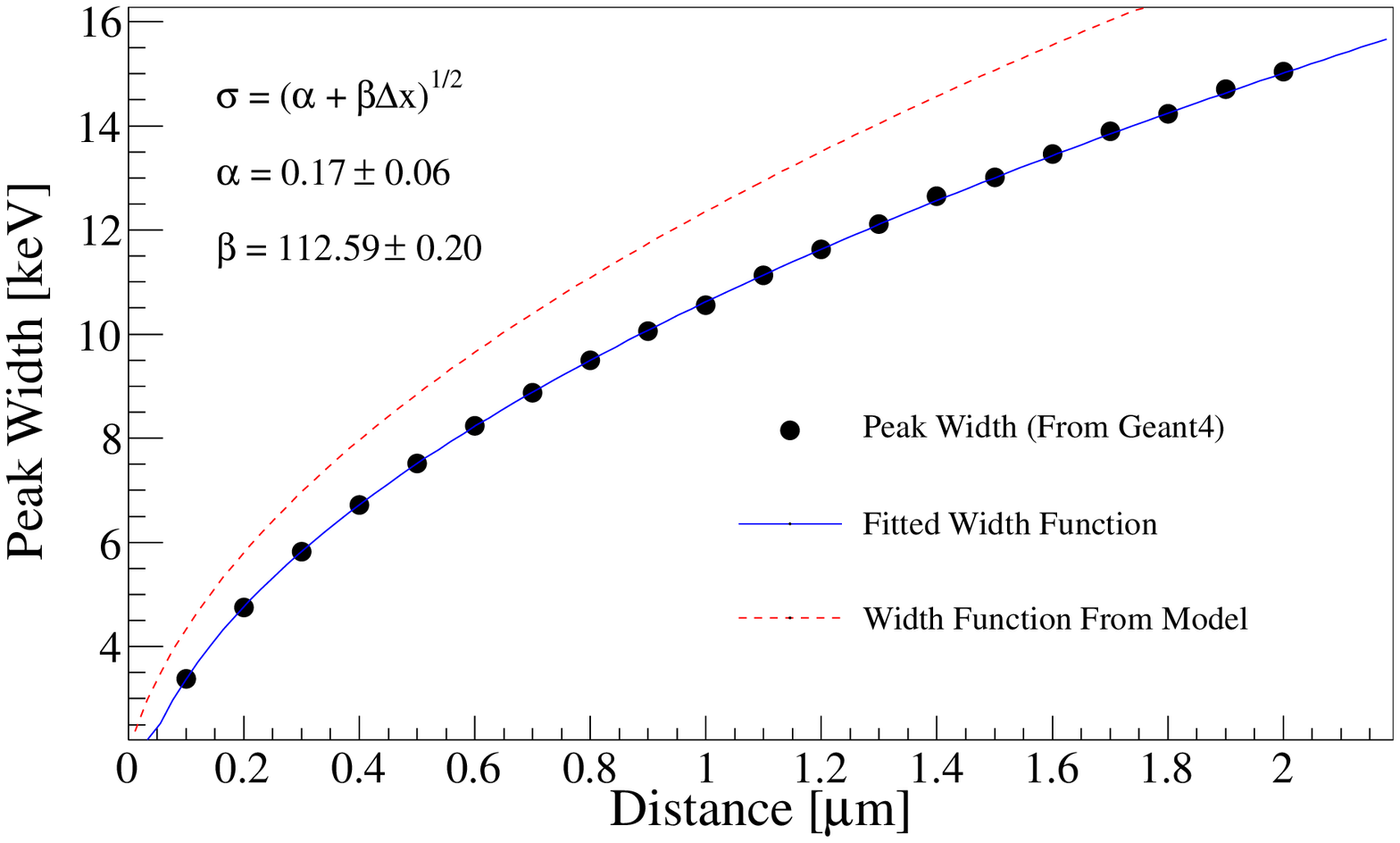}}
\caption{
Discrepancies between simulation and analytic model. When compared to the
measured data, the Geant4 simulation
overestimates the peak position offset (a) while underestimating the peak width
(b). A quadratic function was fit to the peak position offset and a square root
function was fit to the width offset.  These functions were then used to
generate a convolution to modify the simulated energy spectra. The offset
deficit at 0 distance is consistent with the expected offset from nuclear quenching which
is not taken into account in the original Geant4 simulation.}
\label{fig:SimCorrection}
\end{figure}

Simulations of the test stand were also performed using the Geant4- and
ROOT-based software package {\sc MaGe}, with the test stand incorporated into
the package's geometry. A dead layer was simulated in the crystal surface as a
step function, ignoring any energy deposits within the dead region (this
assumption is the same as our analytic treatment). The simulated output spectra
were convolved with a detector-response function, {\em i.e.} a modified gaussian
incorporating the same skew parameter from the model fit.  The width of the
response function was determined by measuring the detector resolution from fits
to gamma peaks in the detector (between 100 and 2600 keV), fitting the
resolution values to a parametrized resolution function $f(E) =
\alpha\sqrt{1+\beta E}$ \cite{Kno00}, and extrapolating the resolution at 5.4
MeV from that function with the fitted values of $\alpha$ and $\beta$.
Comparing the simulations to the data (Fig.  \ref{fig:SimCorrection}) revealed
discrepancies in both the width and offset of the peak structure.  The simulated
offset and width do not match with the data or the analytic model.  A simplified
simulation of a mono-energetic alpha incident on the surface of a detector was
performed for a range of incidence angles (corresponding to a range of dead
layer thicknesses) to determine the difference in alpha peak offsets and widths
between simulation and the model (Fig. \ref{fig:SimCorrection}). These
discrepancies were then used to create a second convolution function that was
applied to the simulated data (as a function of incidence angle). As shown in
Fig. \ref{fig:HoleData}, the corrected simulated spectrum is in good agreement
with the data and the analytic fit.  It should be pointed out that this
particular simulation was performed using Geant4 version 9.0, and simulations
run with newer versions resulted in different discrepancies.

\subsection{External-Bulk Backgrounds}
\label{sec:Bulk}

\begin{figure}[h]
\includegraphics[width=0.95\linewidth]{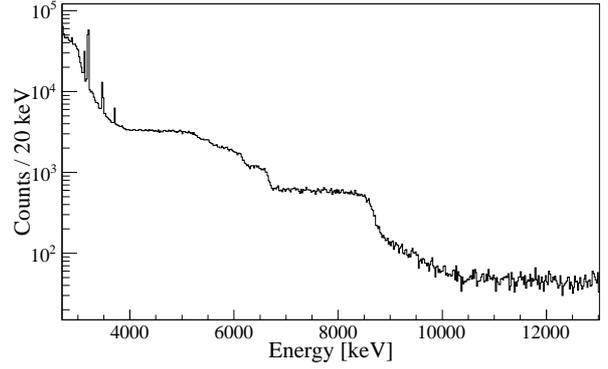}
\caption{\label{fig:ThoriumBareSpectrum} 
Data from a thorium wire source in the test stand.  Alphas from the
\iso{232}{Th} decay chain produce steps in the energy spectrum. The specturm is
shown only from 2.7 MeV to 12 MeV to emphasize the high-energy alpha spectrum.
Data below 2700 keV are dominated by betas and gammas from the \iso{232}{Th}
chain. The peaks visible below 4 MeV are from double- and triple-coincidence
gammas from \iso{208}{Tl}.}
\end{figure}

The rotational feedthrough was replaced with a blank-off for the bulk-type
background study.  A 4 cm length of thorium wire (99\%
\iso{232}{Th})\cite{Goodfellow} was looped 5 mm above the crystal face, allowing
direct shine from the wire to the face of the crystal.  The energy spectrum from
this data set is shown in Fig.  \ref{fig:ThoriumBareSpectrum}.  The close
proximity of the wire to the detector surface means that the detector is
inundated with gammas from the \iso{232}{Th} chain in addition to the alphas.
Double and triple coincidence gammas from \iso{208}{Tl} are evident in the
energy spectrum up to 3710 keV (511 keV + 583 keV + 2615 keV).  The step
structure above 3700 keV is evidence of thorium alphas with the high-energy end
of each step corresponding to the initial kinetic energy of a given alpha.

A model, similar to that of the surface-type detector response, was constructed
to describe bulk-type data on the detector.  The typical range of an alpha in
solid materials of interest (thorium, copper, lead, gold, germanium) is on the
order 10's of $\mu$m.  Thus only the outer portion of material facing the
crystal will contribute alpha backgrounds.  An alpha of original energy $E_0$,
emitted in an external-bulk material at a depth $d$ from the material surface at
an angle $\theta$ with respect to the normal of that surface, will then lose an
average energy similar to Eq. \ref{eq:DeltaE} where $dE/dx$ is a function of the
alpha's energy and distance traveled.  Similarly, energy straggling will widen
the resulting peak. The spectrum from an alpha of initial energy $E_{0}$ at a
depth $d$ emitted at an angle $\theta$ would then be given as

\begin{align}
\label{eq:ThModel}
f\left(E,E_{0},d,\theta\right) &= \frac{1}{\sqrt{2\pi}\sigma} \exp\left(-\frac{E +
E_{0} -\Delta E}{2\sigma^2} \right) \\ \notag
\sigma^2 &=  4\pi N_{a}r^2_{e}(m_{e}c^2)^2\rho\frac{Z}{A}\Delta x \\ \notag
&= 289.9 \Delta x \\ \notag
\Delta E &= \int^{d/\cos\theta}_0 \frac{dE}{dx}dx  .
\end{align}

\noindent where we once again assume a Bohr model of energy straggling and
$\sigma^2$ is calculated using $\rho$, $Z$, and $A$ of thorium.  While $f$ could
be described with an exponentially-modified gaussian as for the surface data,
such details are washed out in practice because of the continuum nature of the
energy spectrum as $f$ is integrated over $\theta$ and $d$.

The model spectrum for a particular decay is then found by integrating Eq.
\ref{eq:ThModel} over $\theta$ and $d$.  Values for $dE/dx$ were taken from
alpha stopping-power and range tables \cite{Astar}, and this integration was
performed for all the alphas in the \iso{232}{Th} decay chain.  The
corresponding energy spectra were then summed, assuming the chain is in secular
equilibrium.  This assumption is valid in our comparison with our data because
all alpha decays in the \iso{232}{Th} chain, except the first, happen within
days of each other.  The first alpha decay in the chain (\iso{232}{Th}) emits a
4 MeV alpha which is not discernable below the beta/gamma continuum in the data
set. 

\begin{figure}[h]
\includegraphics[width=\linewidth]{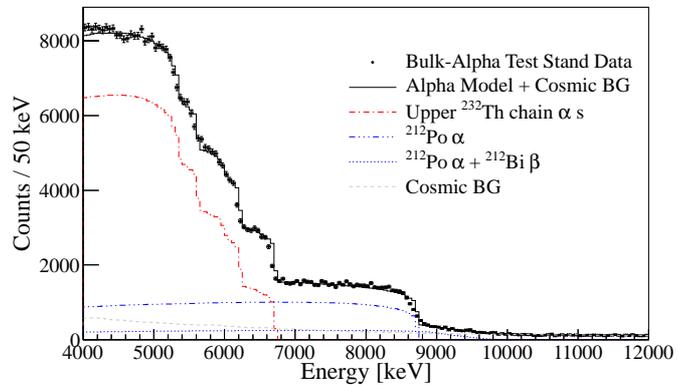}
\caption{Fit of bulk-alpha energy spectrum with composite model from
a thorium-wire source in the test stand.}
\label{fig:ThoriumSpectrum}
\end{figure}

This model was compared with the data set (Fig. \ref{fig:ThoriumSpectrum}).  The
fast alpha decay (300 ns) of \iso{212}{Po}, coming after the beta decay of
\iso{212}{Bi}, occasionally results in pileup in the detector of the alpha (8.8
MeV) and the beta (endpoint 2252 keV, intensity 55.4\%).  To treat this, the
p.d.f. representing the \iso{212}{Po} alpha spectrum was convolved with the beta spectrum of
\iso{212}{Bi}.  A combined p.d.f. consisting of the unmodified \iso{212}{Po} p.d.f., the
\iso{212}{Po}$+$\iso{212}{Bi} p.d.f., the upper \iso{232}{Th} chain p.d.f., and
a cosmic-ray spectrum p.d.f. consisting of a constant plus exponential function
was constructed. This combined p.d.f. was fit to the data set and the results
are shown in Fig. \ref{fig:ThoriumSpectrum}.

[h]

\section{In Situ Measurement}
\label{sec:InSitu}

\begin{figure}[h]
\includegraphics[width=\linewidth]{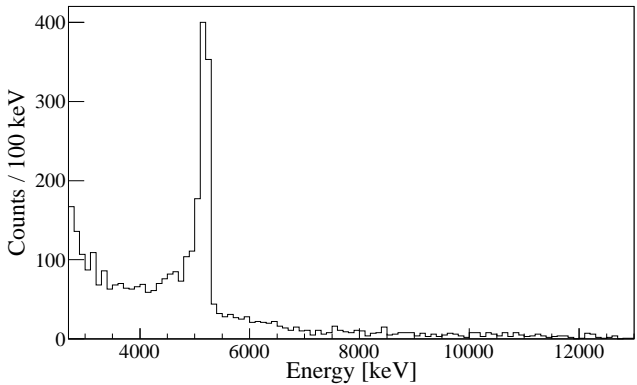}
\caption{\label{fig:InSituSpectrum} Energy spectrum from underground HPGe
detector showing a 5.3 MeV alpha peak (characteristic of the alpha from
\iso{210}{Po} decay).}
\end{figure}

\begin{figure}[h]
\includegraphics[width=\linewidth]{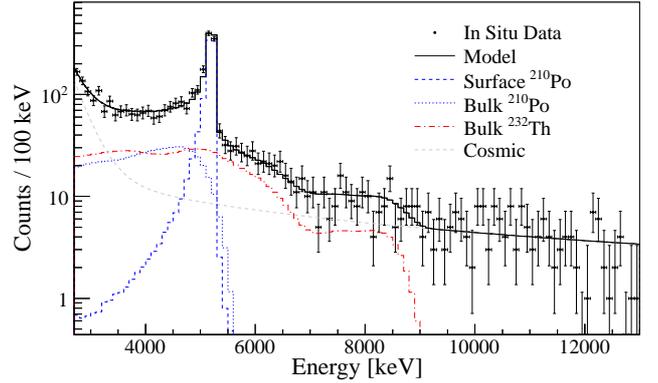}
\caption{\label{fig:InSituFit} Fit to the spectrum of an underground HPGe
detector's energy spectrum with a composite alpha model.  The largest
contributions above 2700 keV are from \iso{210}{Po} surface events and from
external-bulk \iso{232}{Th} and \iso{210}{Po} events.}
\end{figure}

The model developed in the previous section was used to analyze the energy
spectrum of in-situ backgrounds from an existing underground detector.  
An n-type HPGe detector, underground at the Waste Isolation Pilot
Plant near Carlsbad, NM, was discovered to have a peak in its background energy
spectrum at 5.3 MeV as seen in Fig.
\ref{fig:InSituSpectrum} .  We model the spectrum assuming that all events above 2.7
MeV (i.e. above the 2615 keV gamma peak from the decay of \iso{208}{Tl}) must
originate from either surface-alpha events, external-bulk alpha events, or
cosmic events.  A muon-veto scintillator panel was installed above the detector
for a portion of the data taking, giving an energy spectrum of cosmic-ray muons
from events that were tagged with the veto.  The same constant plus exponential
function that was used for the test stand data in Sec.~\ref{sec:TestStand} was
fit to this cosmic-tagged data and subsequently used in the background fit.  A
p.d.f.---consisting of individual p.d.f.s of the surface- and external-bulk
\iso{232}{Th}, \iso{238}{U}, and \iso{210}{Po} events and the cosmic
p.d.f.---was used to fit to the spectrum.  This fit is shown in Fig.
\ref{fig:InSituFit} and the results are summarized in Table
\ref{tab:InSituFitResults}.  Components from the \iso{238}{U} chain---save
\iso{210}{Po}---were consistent with zero and so were excluded from the final
fit.  Alpha yields and rates are tabulated in Table \ref{tab:InSituFitResults}.
The only evidence of bulk events from the \iso{238}{U} chain were found
to be \iso{210}{Po}. This could conceivably come from bulk material containing
short half-life daughters lower in the \iso{238}{U} decay chain (anything below
\iso{226}{Ra}). This would have decayed to the long-lived \iso{210}{Pb} (22 year
half-life), allowing a lead-supported supply of \iso{210}{Po} alphas. 
 
\begin{table}[h]
\centering
\caption {\label{tab:InSituFitResults}Yields and rates of individual
components of composite alpha model, fitted to high-energy data of underground HPGe detector. The data
set represents 317.27 days of livetime.}
\begin{tabular}{lr@{}lr@{}l}\toprule
Contribution & \mc{2}{Yield} & \mc{2}{Counts / Day}  \\  \midrule
Cosmic Events           & 1033&$^{+62}_{-59}$    & $3.33$&$_{-0.19}^{+0.21}$ \vspace{0.05in}\\  
Surface \iso{210}{Po} & 907&$^{+37}_{-36}$     & $2.98$&$_{-0.12}^{+0.12}$ \vspace{0.05in}\\
Bulk \iso{210}{Po}    & 641&$^{+75}_{-76}$     & $1.89$&$_{-0.31}^{+0.31}$ \vspace{0.05in}\\
Bulk \iso{232}{Th}    & 1019&$^{+93}_{-90}$    & $3.27$&$_{-0.60}^{+0.61}$ \vspace{0.05in}\\
\bottomrule
\end{tabular}
\end{table}

\section{Alpha Backgrounds and $\bbdk$}
\label{sec:Impact}

The efficiency for a surface alpha decay on a $p^+$ surface, {\em e.g.} the
alpha from \iso{210}{Po}, to populate the region-of-interest in a double-beta
decay experiment using HPGe detectors was calculated from simulations using
MaGe.  For a nominal dead layer value of 0.3 $\mu$m, the probability that the
energy detected from a \iso{210}{Po} decay on the surface falls within the
$\bbdk$ region-of-interest (ROI) is $(1.21 \pm 0.05)\times 10^{-5}$. Assuming a
possible variation in dead-layer of $\pm 0.01$ $\mu$m adds a systematic
uncertainty of $\pm 0.04\times 10^{-5}$. This includes the corrective
convolution that was required in Section \ref{sec:SurfaceBackgrounds}, although
the difference in calculated efficiency with and without the convolution was
only $0.01 \times 10^{-5}$.   The final, simulated efficiency is then $(1.21 \pm
0.05({\text{stat}}){\pm 0.04({\text{sys}})}) \times 10^{-5}$.  The same efficiency
was also calculated using the surface alpha analytic model. The model was used
to generate an energy spectrum from surface \iso{210}{Po} decays, and the
efficiency for a decay to land in the 4 keV-wide ROI is calculated to be $(1.467
\pm 0.004) \times 10^{-5}$.  Allowing for the same $\pm 0.01$ $\mu$m variation
in dead layer, the model efficiency becomes $(1.47 \pm 0.004({\text{stat}})
{^{+0.04}_{-0.03}} ({\text{sys}}))\times 10^{-5}$.

\begin{figure}[h]
\subfloat[]{\label{fig:DLConsiderI}
\includegraphics[width=0.99\linewidth]{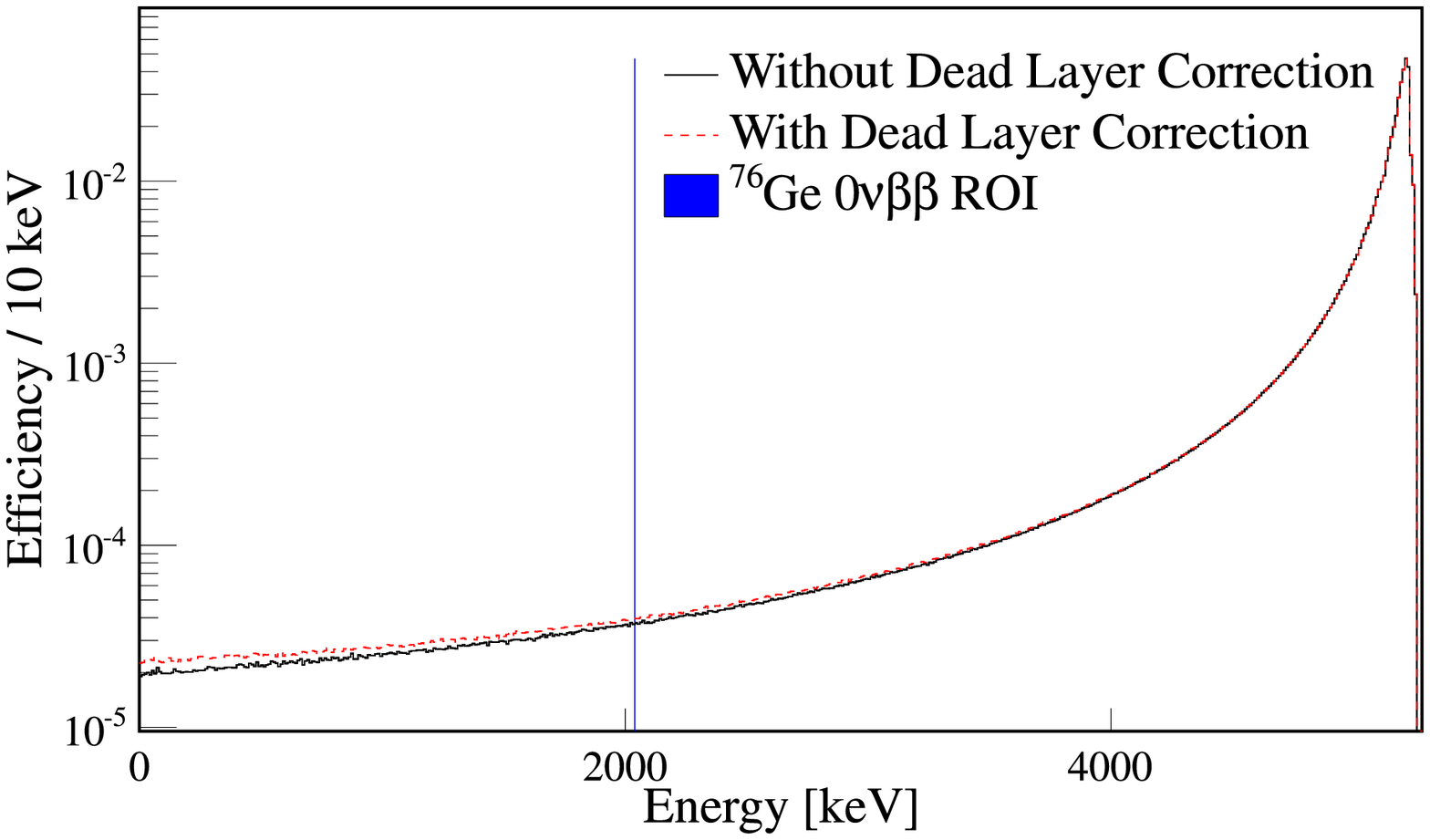}}

\subfloat[]{\label{fig:DLConsiderII}
\includegraphics[width=0.99\linewidth]{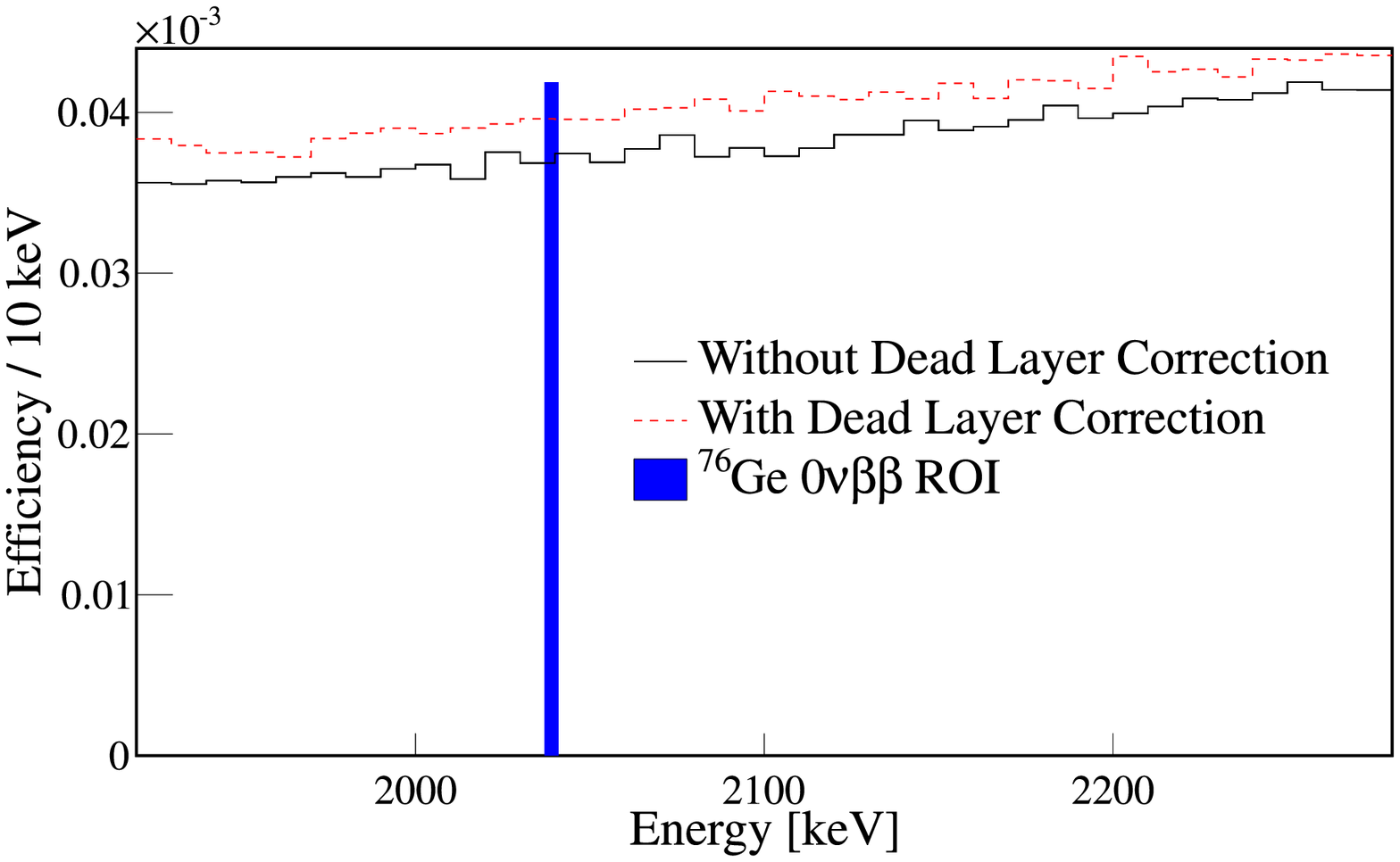}}
\caption[Effect of dead-layer profile on energy spectrum]{The dead-layer profile
(as a function of depth) affects the energy spectrum for surface alphas emitted
from \iso{210}{Po} (5.3 MeV).  The efficiency for surface decays is shown over
(a) the full simulated energy range and (b) a close up around the
region-of-interest at 2039 keV. The solid, black line shows the energy spectrum
from the analytic model with a step-like dead-layer profile.  The red, dashed
line assumes a linear profile. This profile represents the greatest difference
in energy spectra, and the resultant change in efficiency for a surface-alpha
background hit is 7\%.}
\label{fig:DLConsider}
\end{figure}

The two techniques differ by $0.26 \times 10^{-5}$.  The  test-stand data was in
better agreement with the analytic model---particularly the bulk comparison
---and so this value is used as the final calculated efficiency.  The
discrepancy between simulation and model is added as a systematic lower
uncertainty.  Another possible systematic effect arises from uncertainty in the
dead-layer profile.  Both the simulation and the analytic model assume a
step-like efficiency function for the dead layer.  In this case, no charge
collection happens at all within the dead region, but charge collection is 100\%
efficient within the active region.  The stopping power, $\frac{dE}{dX}$, is a
non-linear function of alpha energy, and so the amount of energy lost in a dead
region is dependent upon both energy of the alpha and the charge-collection
efficiency within that region.  A smoothly-varying dead-layer profile will
result in less charge being collected than a step-like profile.  Assuming a
worst-case scenario (with the efficiency profile a linear function, representing
the largest possible difference in the energy spectrum), the energy spectrum can
be compared with the uncorrected (step-function efficiency profile) and this is
shown in Fig.  \ref{fig:DLConsider}.  When this correction is added to the
simulated and model energy spectra, the efficiency for a decay of \iso{210}{Po}
to land in the ROI increases by $7\%$.  A related effect would come from
non-uniformity in the dead layer over the surface of a detector.  If the dead
layer varies from point to point on the detector, then the efficiency for an
alpha decay would vary from point to point as well.  Accounting for the
different predictions of the simulation vs. the analytic model, and folding in
the possibility of a dead-layer effect, the value for the efficiency of a
\iso{210}{Po} alpha, emitted from the $p^+$ (thin) surface of an HPGe detector,
is $(1.47^{+0.10}_{-0.20} \times 10^{-5})$.

\begin{table*}
\centering
\caption[Surface areas for HPGe detector types]
{Surface areas categorized by surface type for a typical n- and p-type detector,
and the modified p-pc natural germanium detectors used in the {\sc Majorana
Demonstrator}. The factor $\lambda$ is the alpha-susceptibility factory as
defined in the text.}
\label{tab:DetSusceptibility}

\begin{tabular}{llr@{ / }rr@{ / }rr@{ / }rrr} \toprule
Detector Type & Mass [kg]  & \mc{7}{Surface Area/Ratio [cm$^2$/\%]} &
\mc{1}{$\lambda$ [cm$^2$/kg]}\\\cmidrule{3-9}
 &  & \mc{2}{passivated} & \mc{2}{$p^+$} & \mc{2}{$n^+$} & Total &  \\ \midrule
n-type &  1.1  &  30.4 & 12.7 & 177.5 & 74.1 & 31.8 & 13.3 & 239.7  & 161.4\\
p-type &  1.1  &  1.8 & 0.8 & 35.4 & 14.8 & 202.6 & 84.5 & 239.8 & 32.2 \\
p-pc   &  0.579  &  1.9 & 1.8 &  0.2 & 0.2 & 104.5 & 98.0 & 106.6 & 0.35\\ \bottomrule
\end{tabular}
\end{table*}

The goals of the {\sc Majorana  Demonstrator} will be to test the $\bbdk$ claim
by Klapdor et al.  \cite{Kla06} and to demonstrate the background goals
necessary for a future tonne-scale experiment.  The alpha background rate of a
detector (or array of detectors), in units of background counts within the
$\bbdk$ ROI per tonne-year, is given as

\begin{equation} 
R_{\alpha} = \frac{k}{M}\int_{S} \mathcal{A}(\vec{r}) \varepsilon
(\vec{r})\Omega(\vec{r})dS, 
\label{eq:alphadecayrate}
\end{equation} 

\noindent where $M$ is the mass of germanium in the detector\footnote[1]{The
usual method is to use $M$ as the amount of active mass, but alpha backgrounds
will not differ for enriched vs. unenriched detectors.}, $\mathcal{A}$ the
surface-alpha activity in Bq/cm$^2$, and $\varepsilon(\vec{r})$ the efficiency
for a decay at position $\vec{r}$ to count as a background.   The integral is
over all relevant surfaces $dS$, and $\Omega(\vec{r})$ is the solid angle of
$p^+$ area with which the area element $dS$ has direct line-of-sight.  The
coefficient $k$ converts the subsequent background rate from counts per
second-kilogram into counts per tonne-year.  This formula can be significantly
simplified for decays that occur on the surface of the detector itself:

\begin{equation} 
R_{\alpha} = k\frac{\mathcal{S}}{M}\mathcal{A}_{avg}\varepsilon = k \lambda
\mathcal{A}_{avg}\varepsilon
\label{eq:simplifiedalpharate}
\end{equation} 

\noindent with $\mathcal{S}$ the total susceptible surface area,
$\mathcal{A}_{avg}$ the average surface-activity rate, and $\varepsilon$ the
efficiency calculated in the previous section (a true ``surface'' event with
solid angle $\Omega = 2\pi$).  The factor $\lambda$ is the ratio of susceptible
surface area to active mass (with units of area/mass).  Because the {\sc
Demonstrator} will be composed of modified, p-type point contact (p-pc)
detectors, it will have a particularly-low susceptibility to alpha backgrounds,
especially in comparison to n-type detectors (Table
\ref{tab:DetSusceptibility}).  This is because the vast majority of surface area
that is $n^{+}$, or ``thick''.  The susceptibility for the modified p-pc
detectors expected to be used in the {\sc Demonstrator} is 0.34$\pm 0.03$
cm$^{2}$/kg, where the uncertainty comes from the quoted tolerances of the
detector dimensions for the Canberra BEGe \iso{nat}{Ge} detectors that will be
installed in the first module of the {\sc Majorana Demonstrator}. For
comparison, the susceptibility for a typical n-type detector is over 400 times
greater.  Plugging in the susceptibility and the calculated efficiency, the rate
then becomes 

\begin{align}
R_{\alpha} &= 1.57^{+0.16}_{-0.17} \times 10^{5} \mathcal{A}_{avg}, \qquad
\qquad \mathcal{A}_{avg}\text{ in Bq/cm}^2 \notag \\ &= 1.82^{+0.20}_{-0.21}
\mathcal{A}_{avg} \qquad \qquad  \mathcal{A}_{avg}\text{ in Decays/Day/cm}^2
\label{eq:PPCRate} 
\end{align}

\noindent Table \ref{tab:BGAlphaRates} displays expected background rates from
\iso{210}{Po} alphas for several values of surface activity $\mathcal{A}_{avg}$.
The alpha rate from the Sudbury Neutrino Observatory's \iso{3}{He} proportaional
counter detectors represents the cleanest detector surface ever measured in
terms of alpha contamination\cite{Ams07}.  Also shown are the rates for a
typical n-type detector, given the same activity rates.

If the alpha decays are not coming from the surface of the HPGe crystal, then
the solid angle simplification no longer applies.  For configurations involving
complicated surfaces facing the $p^+$ area of a detector, the integral in Eq.
\ref{eq:alphadecayrate} can be calculated using Monte Carlo methods.  An upper
limit on the activity can also be placed, assuming that the $p^+$ region of the
crystal has a line-of-sight view of a contaminated surface with average surface
activity $\mathcal{A}_{avg,ext}$.  The limit is then

\begin{equation}
R_{\alpha, ext} \le k \lambda \mathcal{A}_{avg,ext}\varepsilon
\label{eq:generalalpharate}
\end{equation}

\noindent where the equality holds if all of the solid angle that the sensitive
area ``sees'' is emitting alphas at the surface rate $\mathcal{A}_{avg,ext}$
(and Eq. \ref{eq:simplifiedalpharate} is the limiting case for this).  Equation
\ref{eq:generalalpharate} holds via a simple flux argument.

\begin{table}[h]
\centering 
\caption[Estimated background count rates from surfacealphas] 
{Estimated background count rates from surface alphas for modified p-pc
detectors as will be used in the {\sc Majorana Demonstrator}, and for typical
n-type detectors.  These rates are based upon assuming various surface-activity
levels.  The clean room entry corresponds to a class-100 clean room built for
radon-deposition testing for the Borexino experiment \cite{Leu06}, and
surface-activity values in the table represent the radon-deposited activity on
nylon in that clean room.  The calculation for the SNO \iso{3}{He} detectors is
found in \cite{Sto05}.  
}
\label{tab:BGAlphaRates}
\begin{tabular}{llll}\toprule
Source & Activity & \mc{2}{Background Rate} \\
  &  [Bq/cm$^2$] & \mc{2}{[Counts in ROI / t-y]} \\\cmidrule(r){3-4}
& & {\sc p-pc}\qquad \qquad \qquad  &  n-type \\
\midrule
Clean Room& $1.0 \times 10^{-6}$ & 0.16 & 70\\
In Situ Detector & $9.0 \times 10^{-7}$ &  0.14 & 61\\
MJ BG Model& $5.0 \times 10^{-7}$ &  0.08 & 35\\
SNO \iso{3}{He} Detector & $5.0 \times 10^{-9}$ & 0.0008 & 0.35\\
\bottomrule
\end{tabular}
\end{table}

The susceptible surface of a modified p-pc detector, located right at the point
contact, only has a direct line-of-sight with the detector mount.  Alphas can
only pose a background if they originate from surface plate-out (on the surface
of the detector or the detector mount) or from the bulk material of the detector
mount.  Because of this compartmentalization, the rate of alpha backgrounds (in
counts per tonne-year) should be the same regardless of the number of detectors.
The rate formula for p-pcs (Eq.  \ref{eq:PPCRate}) would still be valid, then,
for a one-tonne scale experiment made up of p-pc detectors with the same
susceptibility factor $\lambda$.  The usage of p-pc detectors is extremely
beneficial to {\sc Majorana} from a surface-alpha standpoint (to say nothing of
their other important qualities).  As Table \ref{tab:BGAlphaRates} notes, the
difference in surface-alpha background rates between p-pc detectors and n-types
is large.  Without further cuts from pulse-shape analysis, Table
\ref{tab:BGAlphaRates} makes it clear that n-type detectors are unsuitable for
$\bbdk$ experiments without heroic measures to limit surface activity. A final
comment should be made about the passivated surface--the annulus of material
between the $n^+$ and $p^+$ surfaces. This area has 10 times the surface of the
$p^+$ layer for the modified p-pc detectors, and so must be addressed. Our test
stand was modified to allow the \iso{241}{Am} source to have a direct shine path
on the passivated surface of the crystal. Not only were no alphas present
visible in the resultant energy spectrum, but no 59 keV gammas were visible
either. If the passivated surface were as susceptible to alphas as the $p^+$
surface, the background count rate from a given surface contamination would
increase by a factor of 10, but still be far better than that from n-type
detectors. We are continuing to study the issue.

\section{Conclusion}
\label{sec:Conclusions}

We built a test stand for the purpose of studying surface-alpha decays of
radioactive isotopes on surfaces of HPGe detectors.  A simple detector response
model was constructed and fit to test-stand data, and Geant4 simulations
were compared with test stand data and the detector response model.  The
efficacy of simulations of surface-alpha decays for $\bbdk$
experiments was examined and we propose a correction factor to be used for such
simulations.  We calculate allowable contamination values for n-type and p-pc
HPGe detectors used in double-beta decay searches given desired surface-alpha
background assumptions.

\section*{Acknowledgements}
We gratefully acknowledge support from the Office of Nuclear Physics in the U.S. Department of
Energy (DOE) Office of Science under grant numbers
DE-FG02-97ER4104, DE- FG02-97ER41033, and DE-FG02-97ER41020 as well as
contract number 2011LANLE9BW.
We also acknowledge the support of the DOE through the 
the LANL/LDRD program. Finally, we thank our friends and
hosts at the Waste Isolation Pilot Plant (WIPP) for their continuing support of
our activities underground at that facility.  


\bibliography{AlphaPaper}

\end{document}